\documentclass[twocolumn,superscriptaddress,secnumarabic,amssymb, nobibnotes, aps, pra]{revtex4-2}
\usepackage[utf8]{inputenc}
\usepackage{graphicx}
\usepackage{placeins}
\usepackage{floatflt}
\usepackage{epsfig,epsf,rotating}
\usepackage{amsmath}
\usepackage{amsopn}
\usepackage{amssymb}
\usepackage{bm}
\usepackage[normalem]{ulem}
\usepackage{booktabs, makecell}
\usepackage{comment}

\begin{document}
	
	\title{Photoassociative Spectroscopy of $^{87}$Sr}
	\author{J.C. Hill}
		\affiliation{Department of Physics and Astronomy, Rice University, Houston, TX 77005, USA}
        \affiliation{Applied Physics Graduate Program, Smalley-Curl Institute, Rice University, Houston, TX 77005, USA}
	\author{W. Huie}
    \author{P. Lunia}
    \author{J.\,D. Whalen}
    \author{S.\,K.\,Kanungo} 
    \author{Y. Lu}
    	\affiliation{Department of Physics and Astronomy, Rice University, Houston, TX 77005, USA}
	\author{T.C. Killian}
	\email[]{killian@rice.edu}
	    \affiliation{Department of Physics and Astronomy, Rice University, Houston, TX 77005, USA}
	  \affiliation{Applied Physics Graduate Program, Smalley-Curl Institute, Rice University, Houston, TX 77005, USA}

	\date{\today}
	
\begin{abstract}
We demonstrate photoassociation (PA)  of ultracold fermionic $^{87}$Sr atoms.  The binding energies of a series of molecular states on the $^1\Sigma^+_u$ $5s^2\,^1$S$_0+5s5p\,^1$P$_1$ molecular potential are fit with the semiclassical LeRoy-Bernstein model, and 
 PA resonance strengths are compared to predictions based on the known $^1$S$_0+^1$S$_0$ ground state potential. Similar measurements and analysis were performed for the bosonic isotopes $^{84}$Sr and $^{86}$Sr, allowing a combined analysis of the long-range portion of the excited-state potential and determination of the $5s5p\,^1$P$_1$ atomic state lifetime of $5.20 \pm 0.02$\,ns. The results enable prediction of PA rates across a wide range of experimental conditions. 
\end{abstract}

	\maketitle

\label{sec:Introduction}

Ultracold gases of the various isotopes of strontium are currently studied for a wide range of applications, such as frequency metrology \cite{chm17,nye19,bhh19}, quantum simulation of many-body physics \cite{ghg10,mbs13}, quantum information \cite{mcs20,hps20}, Rydberg physics \cite{dky16,csw18}, and cold collisions \cite{mmp08,tcm09,ydr13,ahd18}. The fermionic isotope ($^{87}$Sr) in particular attracts significant attention because it is used in optical atomic clocks \cite{lby15,chm17} and the ground-state tenfold degeneracy arising from the large nuclear spin ($I=9/2$) introduces novel magnetic phenomena \cite{ghg10}.
For many of these experiments, ultracold samples are trapped in optical lattices in order to prevent atom-atom interactions or introduce spatial periodicity for quantum simulation of materials. Photoassociative (PA) spectroscopy \cite{wbz99,jtl06}, which is the optical formation of bound molecules from the initial state of two colliding atoms, is a well-established and useful technique for probing ultracold gases in optical lattices. It can be used to detect double occupancy of lattice sites \cite{rbm04}, which provides a measurement of sample temperature and probe of quantum phase transitions \cite{tys12}. This technique has not been utilized for $^{87}$Sr, however, and no study on PA in this isotope has been published.  
PA formation of molecules bound in the $5s^2\,^1$S$_0$+$5s5p\,^3$P$_1$ molecular potential in $^{87}$Sr was mentioned in \cite{Sonderhouse2020}, and this is more challenging than in bosonic isotopes \cite{zbl06,bmc14,rrm18} because of the large nuclear spin and hyperfine splitting of the molecular states. 

Here, we report photoassociation in $^{87}$Sr to states on the $^1\Sigma^+_u$ $5s^2\,^1$S$_0+5s5p\,^1$P$_1$ molecular potential, to the red of the principal transition at $\lambda=460.85$\,nm. Hyperfine splitting in the excited state is small ($\sim 60$\,MHz) and unresolved, producing a simple spectrum. We report line strengths in terms of PA-collision-event rate constants for transitions across a wide range of binding energies and provide parameters for a fit of the binding energies to the semi-classical Leroy-Bernstein formula \cite{lbe70}. This can inform design of experiments using PA spectroscopy as a probe of $^{87}$Sr in optical lattices.
We also report Leroy-Bernstein parameters describing molecular binding energies for $^{84}$Sr and $^{86}$Sr. 
The extracted value of the $C_3$ coefficient and associated $^1$P$_1$ atomic lifetime are compared with previous results from PA spectroscopy of $^{88}$Sr \cite{nms05,ykt06} and $^{86}$Sr \cite{mms05} and measurements of AC Stark shifts \cite{hps20}. PA-collision-event rate constants are compared to predictions \cite{bav00} based on the known $^1\Sigma^+_g$ $5s^2\,^1$S$_0$+$5s^2\,^1$S$_0$ potential \cite{mmp08,skt10}.

\label{sec:ExpMeth}
Using methods described in \cite{ssk14}, ultracold strontium atoms are trapped in an optical dipole trap ($1064$\,nm wavelength) consisting of two crossed elliptical beams  propagating perpendicular to gravity with tight axis along gravity. Atoms are initially loaded into the trap and then a short stage of evaporation produces the sample used for PA spectroscopy. For $^{87}$Sr, spectroscopy is performed in a trap with oscillation frequencies ($f_x,f_y,f_z$)=($68,85,433$)\,Hz. At the start of PA laser exposure,the number of atoms is $N=2.3\times 10^6$, the temperature is $T=1.8\,\mu$K, and the peak density is $n_0=2.1\times 10^{13}$\,cm$^{-3}$. The sample has approximately equal population of the $10$ nuclear spin states \cite{wkd19}. For $^{84}$Sr, these parameters are ($f_x,f_y,f_z$)=($81,30,406$)\,Hz, $N=4.5\times 10^6$, $T=1.9\,\mu$K, and $n_0=1.7\times 10^{13}$\,cm$^{-3}$. For $^{86}$Sr, these parameters are ($f_x,f_y,f_z$)=($61,76,388$)\,Hz, $N=1.2\times 10^6$, $T=1.8\,\mu$K, and $n_0=1.1\times 10^{13}$\,cm$^{-3}$. The PA laser intensity ($3.6-237$\,mW/cm$^2$) and exposure time ($10-1000$\,ms) are varied depending upon the sample and strength of the PA transition. Typical peak atom loss  due to PA is 10-50\%. The sample temperature varies by no more than 25\% for measurements of the PA rate constant. The PA laser beam has $e^{-2}$ radii of $w_{\textrm{horz}}=850\,\mu$m and $w_{\textrm{vert}}=440\,\mu$m on the atoms, and it is treated as homogeneous over the sample.

\begin{figure*}
	\includegraphics[width=17.2cm]{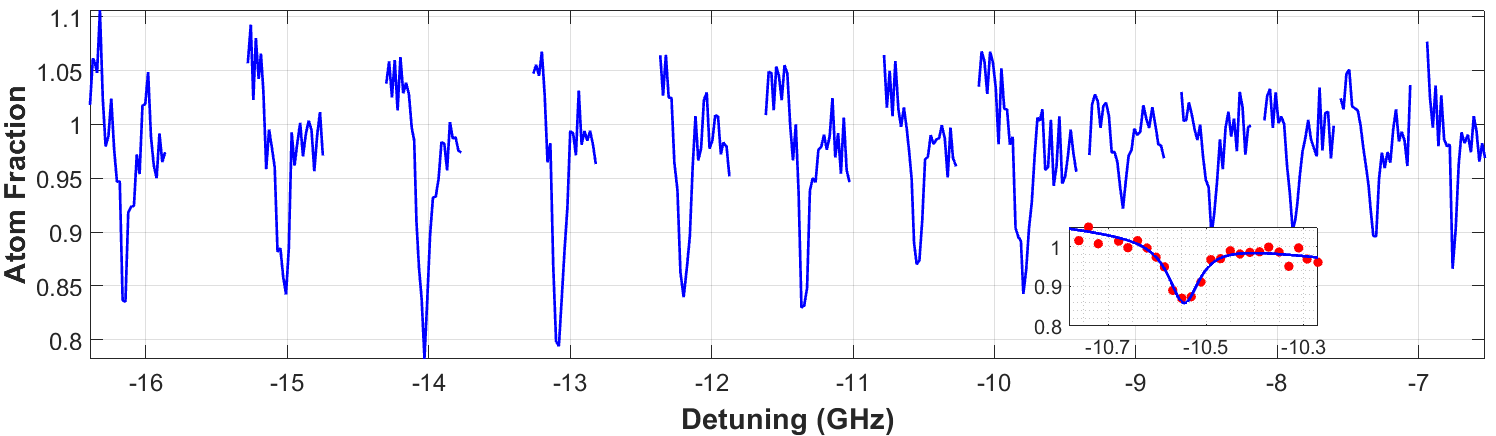}
	\caption{Representative atom-loss spectra for  excitation to molecular states on the  $^1\Sigma^+_u$ $5s^2\,^1$S$_0+5s5p\,^1$P$_1$ molecular potential in $^{87}$Sr. Background atom number is normalized to one. PA laser exposure times and intensities are adjusted to give comparable atom loss for all spectra, so line intensity does not reflect the PA-collision-event rate constant. Exposure times and intensities vary from 75-200ms and 3.6-11.5 mW/cm$^{2}$ respectively. Inset is a representative fit to a Lorentzian.
	}
	\label{fig:samplespectra}
\end{figure*}

After exposure, the PA laser and the dipole trap are extinguished, and the atom number and sample temperature are measured with resonant time-of-flight absorption imaging on the 461\,nm transition. Photoassociation is indicated by atom loss. Figure \ref{fig:samplespectra} shows representative atom-loss spectra for  $^{87}$Sr. The PA laser is locked to a wavemeter (Moglabs FZW600) that is calibrated against the atomic $5s^2\,^1$S$_0\rightarrow 5s5p\,^1$P$_1$ transition, with a frequency accuracy of $\sim\pm 30$\,MHz.


The PA resonances, labeled by vibrational index $\nu$, are fit to a Lorentzian lineshape. Typical linewidths range between $60-150$\,MHz. This is consistent with the lower limit given by twice the atomic linewidth $\gamma_{atomic}=30.24$\,MHz \cite{ykt06}. The dominant source of additional line broadening is laser frequency jitter. Binding energies (defined as $E_{\nu} >0$) are determined by taking the difference between PA resonance positions and the wavenumber corresponding to excitation of the $5s^2\,^1$S$_0\rightarrow 5s5p\,^1$P$_1$  atomic resonance. 

The long-range form of the excited-state $^1\Sigma^+_u$ $^1$S$_0$+$^1$P$_1$ molecular potential can be approximated as
\begin{equation}\label{Eq:potential}
  V_{e}(r)=D-\frac{C_3}{r^3}+\frac{\hbar^2[J(J+1)+2]}{2\mu\, r^2}, \quad C_3=\frac{3\hbar \lambda^3}{16 \pi^3 \tau}
\end{equation}
where D is the dissociation energy, $\mu$ the reduced mass, $r$ is the internuclear separation, and $\tau=1/(2\pi\gamma_{atomic})$ is the lifetime of the  $^1$P$_1$ atomic state.
Because of the ultracold temperature, only $s$-wave collisions occur and only $J=1$ molecular rotational states are excited. The rotational energy is small and can be neglected, as can thermal energy of the initial collisional state. The binding energies, $E_{\nu}$, can be fitted to the semi-classical Leroy-Bernstein formula \cite{lbe70}.
\begin{equation}
\label{Eq:LB}
E_{\nu} =[(\nu-\nu_{D})H_3]^6, \quad H_3 = \frac{1}{C_3^{\frac{1}{3}}}\frac{\hbar\Gamma\left(\frac{4}{3}\right)}{2\Gamma\left(\frac{5}{6}\right)}\sqrt{\frac{2\pi}{\mu}}
\end{equation}
where $\nu=1$ corresponds to the least-bound  state, $\Gamma$ is the gamma function, and $\nu_D$ is a fit parameter ranging from 0 to 1 that takes on a different value for each isotope. We perform a combined fit of all data to Eq. \ref{Eq:LB} using a common value of $C_3$ as a fit parameter and independent values of $\nu_D$ for each isotope. The quantum numbers $\nu$ assigned to each level are shifted to obtain $0<\nu_D<1$.

\begin{figure}[!h]
	\centering
	\includegraphics[width=8.6cm]{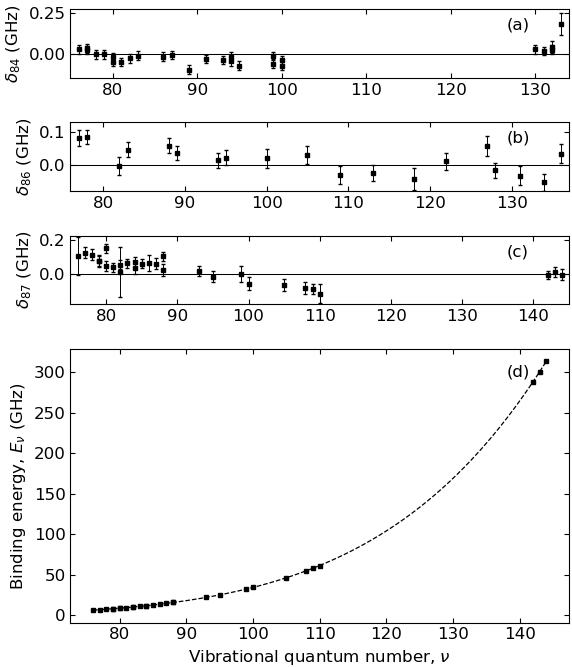}
	\caption[title]{Fit of the binding energies of states on the  $^1\Sigma^+_u$ $5s^2\,^1$S$_0+5s5p\,^1$P$_1$ molecular potential for various strontium isotopes. Plots 
(a-c) show the difference ($\delta$) between observed and fit values for the binding energies for $^{84}$Sr, $^{86}$Sr, and $^{87}$Sr respectively, using parameters from Tab.\ \ref{Table:FitParameters}.
Plot (d) shows the measured binding energies as a function of vibrational quantum number, $\nu$,  for $^{87}$Sr. The dashed line is the best fit of the data to Eq.\ \ref{Eq:LB}. }
	\label{fig:BindingEnergiesAndFits}
\end{figure}

Figure \ref{fig:BindingEnergiesAndFits} shows binding energies for $^{87}$Sr and residuals for all isotopes. Table \ref{Table:FitParameters} shows the fit parameters, including the value of $\tau$ extracted from the $C_3$ coefficient. The statistical uncertainty in $\tau$ from this procedure is very small, but the fit residuals for all three isotopes show a systematic trend corresponding to $\sim 100$\,MHz variation over a change of $\sim 100$\,GHz in binding energy. This trend might indicate systematic wavemeter error, variation in molecular-state AC Stark shifts from the optical dipole trap laser fields, or the influence of additional terms in the molecular potential not accounted for by  the Leroy-Bernstein formula \cite{com04}. Adding an additional fit parameter in the form of a binding energy offset removes the systematic trends in the residuals and increases $\tau$ by $0.02$\,ns or 0.4\%, which we take as our uncertainty. We thus quote a final value of $\tau=5.20\pm 0.02$\,ns. The value of $\tau$ determined here differs by 1\% from the most accurate reported measurement of $\tau=5.263\pm 0.004$\,ns, performed with photoassociation in an optical lattice \cite{ykt06}, and is in closer agreement with a more recent value of $\tau=5.234\pm 0.008$\,ns determined from AC Stark shifts of Sr levels \cite{hps20}. The binding energies for all observed transitions are included in the supplementary material. 

\begin{table}[h!]
  \begin{center} \setcellgapes{3pt}\makegapedcells
    \begin{tabular}{|c|c|c|c|}
      \hline 
        \textrm{Isotope} & $^{87}$Sr & $^{86}$Sr & $^{84}$Sr  \\
      \hline
      	$\nu_D$   & 0.87 $\pm$ 0.02 & 0.20 $\pm$ 0.02 & 0.87 $\pm$ 0.02 \\
      \hline
       $\tau$ (ns) & \multicolumn{3}{c|}{5.20 $\pm$ 0.02}  \\
       \hline 
    \end{tabular}
      \caption{Parameters from a fit of binding energy for $^{84}$Sr, $^{86}$Sr, and $^{87}$Sr  to  Eq.\ref{Eq:LB}.}
      \label{Table:FitParameters}
  \end{center}
\end{table}

 The loss of atoms due to photoassociation is described by a local equation for
the evolution of the atomic density
\begin{equation}\label{densitydecay}
\dot{n}=-\beta (I,f) n^2-\Gamma_{1} n,
\end{equation}
where $\Gamma_{1}$ describes one-body loss due to light scattering on the atomic transition or background gas collisions, and the two-body loss is described by $\beta$, which depends on PA laser intensity, $I$, and frequency, $f$. Assuming constant sample temperature, the number of atoms in the trap as a function of the PA exposure time $t$ is given by
\begin{equation}\label{number}
   N(t)={N_0 \rm{e}^{-\Gamma_{1} t} \over 1+
   {n_0\beta \over 2\sqrt{2}\Gamma_{1}}(1-\rm{e}^{-\Gamma_{1} t})},
\end{equation}
where  $N_0$  is the number at the beginning of the PA interaction
time and $n_0$ is the initial density. The density can be calculated from the number and sample
temperature, approximating the trap as an infinitely deep harmonic potential with oscillation frequencies matching the measured frequencies of the optical trap. Eq. 3. has been integrated over volume to yield the evolution of sample number, which is solved by Eq. 4.

Near resonance with the transition to a molecular state with vibration quantum number $\nu$ and center frequency $f_{\nu}$, the PA loss is described with a Lorentzian lineshape \cite{mms05},
\begin{equation}\label{beta}
  \beta=\frac{2 K_{\nu} \gamma_{mol}}{\gamma}\frac{1}{1+4(f-f_{\nu})^2/\gamma^2},
\end{equation}
where $\gamma_{mol}=2\gamma_{atomic}$ is the natural linewidth of the PA transition due to radiative decay of the molecular state and $\gamma$ is the observed linewidth of the transition. $K_{\nu}$, which is proportional  to laser intensity $I$,  is the resonant collision-event rate constant that would be observed  in the absence of any broadening beyond the natural linewidth. Thermal broadening is much smaller than $\gamma_{mol}$  and the laser intensity is low enough that saturation effects are negligible.

\begin{figure}[!t]
	\centering
	\includegraphics[width=8.6cm]{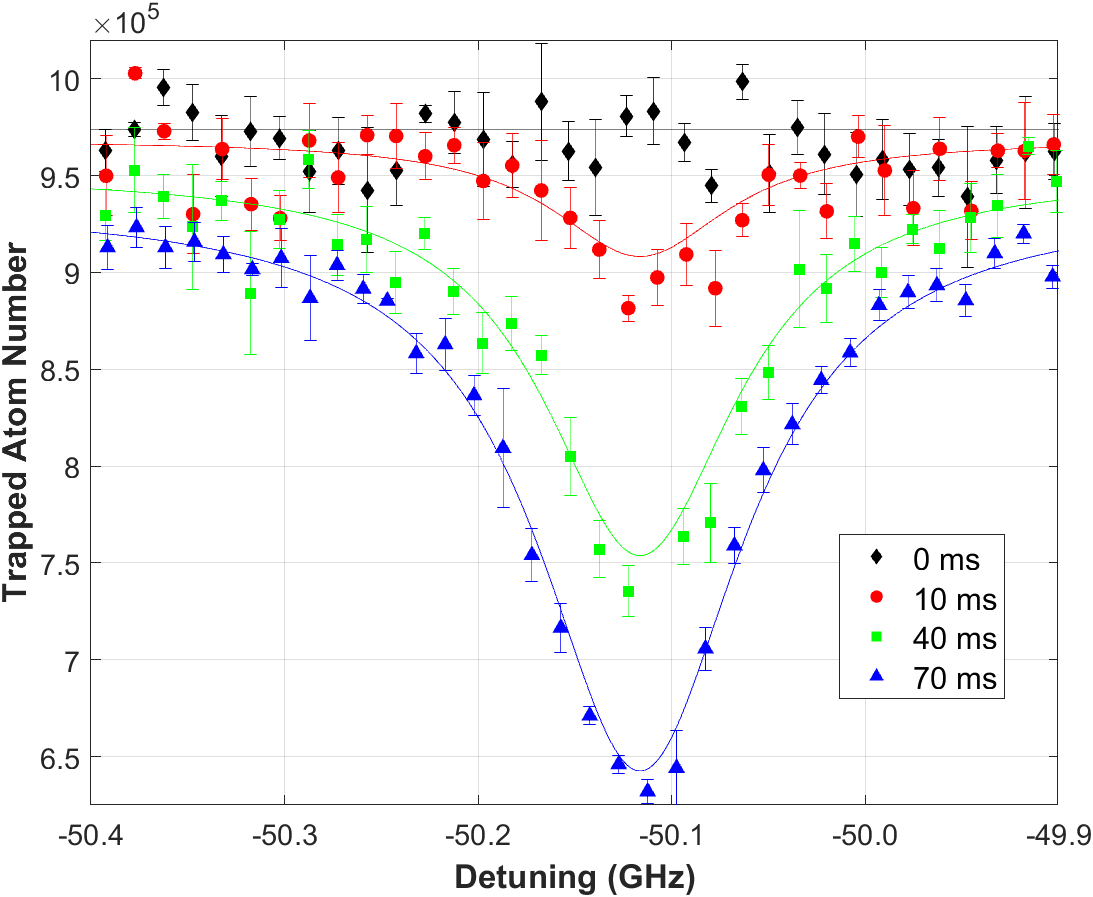}
	\caption[title]{Atom loss spectra for PA to the $\nu=105$  state of the $^1\Sigma^+_u$ $5s^2\,^1$S$_0+5s5p\,^1$P$_1$ molecular potential for $^{86}$Sr.
 PA laser exposure time is indicated in the legend. A combined fit of all data to Eqs.\ \ref{number} and \ref{beta} determines the
 the peak PA collision-event rate constant $K_{\nu}$. The PA laser intensity is 3.6\,mW/cm$^2$ and the initial density is $n_0=1.3 \times 10^{13}$\,cm$^{-3}$. The fit result is $K_{\nu}/I=2.8\times 10^{13}$\,cm$^{5}$/(s$\cdot$mW).}
	\label{fig:FitLossSpectra}
\end{figure}

Figure \ref{fig:FitLossSpectra} shows a fit of  typical atom-loss spectra to Eqs.\ \ref{number} and \ref{beta} for different PA laser exposure times for $^{86}$Sr.  All fit parameters are well determined, but systematic uncertainty in laser intensity, atom density, and sample temperature lead to systematic uncertainty of about a factor of 3 in
the  fundamental quantity for comparison with theory, $K_{\nu}/I$.

$K_{\nu}/I$ is proportional to $F^{\nu}_{eg}(E,\Delta_{\nu})$, the free-bound Franck-Condon factor for excitation to the excited state $\nu$  from the ground $^1\Sigma^+_g$ $5s^2\,^1$S$_0$+$5s^2\,^1$S$_0$ potential, where $E$ is the initial collision energy, and
$\Delta_{\nu}<0$ is the detuning from atomic resonance. Through the relation 
\begin{equation}\label{Eq:DetuningAtCondonPoint}
 h\Delta_{\nu}=V_e(r^{\nu}_{C})-D-V_g(r^{\nu}_{C})
\end{equation}
the detuning defines the Condon radius, $r^{\nu}_{C}$, which is the internuclear separation at which the photon energy (via the Plank constant $h$) is resonant with the difference in molecular potentials. $r^{\nu}_{C}$ can be interpreted as the classical separation at which excitation occurs.
At long range and ultracold temperatures, the ground potential can be approximated by $ V_{g}(r)=-{C_6}/{r^6}-{C_8}/{r^8}-{C_{10}}/{r^{10}}$.
 
Using the reflection approximation \cite{bav00}, the Franck-Condon factor can be related to the energy-normalized ground-state wave function at the Condon point, $\Psi_g(r^{\nu}_C,E)$ through
\begin{equation}\label{Eq:FCFactor}
 F^{\nu}_{eg}(E,\Delta_{\nu})=\frac{\partial E_{\nu}}{\partial \nu}\frac{1}{d_c}|\Psi_g(r^{\nu}_C,E)|^2
\end{equation}
where ${\partial E_{\nu}}/{\partial \nu}$ is the spacing between adjacent vibrational levels in the excited state at level $\nu$,  which can be found from Eq.\ \ref{Eq:LB}. Also,
\begin{equation}\label{Eq:slopedifference}
  d_c=\left|\frac{d}{dr}\left[V_e(r)-V_g(r) \right] \right|_{r=r^{\nu}_{C}}
\end{equation}
is the difference in slopes of the excited and ground potentials.

We calculate ground-state wavefunctions for a collision energy of $2$\,$\mu$K by numerically integrating the Schr{\"o}dinger equation \cite{joh77}. We use the full ground-state potential from \cite{skt10} in the form from \cite{skt08} with recommended values of $C_6=1.525\times 10^{7}\,\mathrm{cm}^{-1}{\mathrm{\AA}}^6$, $C_8=5.159\times 10^8 \mathrm{cm}^{-1}{\mathrm{\AA}}^8$, and $C_{10}=1.91\times 10^{10}\,\mathrm{cm}^{-1}{\mathrm{\AA}}^{10}$.
\begin{figure}[!t]
	\centering
	\includegraphics[width=8.6cm]{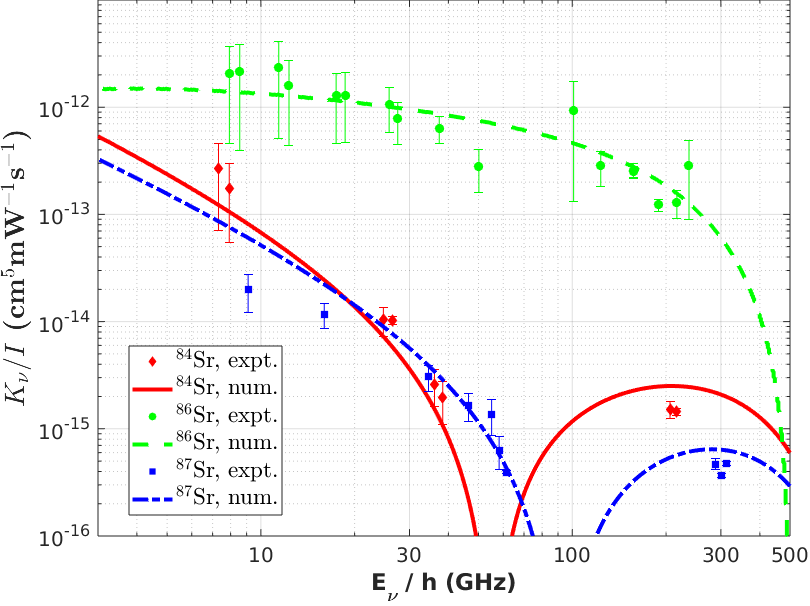}
	\caption[title]{ Intensity-normalized resonant collision-event rate constant ($K_{\nu}$) for strontium PA transitions. Solid lines are predictions based on Franck-Condon factors derived from numerically calculated ground state wave functions (Eq.\ \ref{Eq:FCFactor}) as described in the text.}
	\label{fig:KValues}
\end{figure}
Figure \ref{fig:KValues} shows the measured values of $K_{\nu}/I$ and theoretically expected values  based on the calculated wavefunctions.
A common scaling factor is applied to theoretical predictions to account for the proportionality between $K_{\nu}/I$ and the Franck-Condon factors. Predicted rates for $^{87}$Sr are further reduced with respect to the bosons by the ratio of the pair correlation functions $g^{(2)}_f/g^{(2)}_b$, where  $g^{(2)}_b=2$ reflects bunching for bosons and $g^{(2)}_f=0.9$ reflects Pauli exclusion for a gas of identical fermions in ten equally populated internal states \cite{wkd19}. Measurements for $^{86}$Sr agree within a factor of two with previously reported values \cite{mms05}, which is reasonable given systematic uncertainties.
 

The rate coefficients go to zero when the Condon radius for the transition is near a node of the ground-state wave function (Fig.\ \ref{fig:KValues}). For  $^{87}$Sr and $^{84}$Sr, the node interrogated by measurements reported here corresponds to internuclear spacing equal to the $s$-wave scattering lengths, $a_{84}=123$\,$a_0$ or $a_{87}=96$\,$a_0$  \cite{mmp08}, where $a_0$ is the Bohr radius. For $^{86}$Sr, $a_{86}=823$\,$a_0$ is a larger lengthscale than probed here, and the node corresponding to $\Delta=-500$\,GHz is the second pre-asymptotic node \cite{mms05}. 


For $^{87}$Sr and $^{84}$Sr, the two isotopes most often used for quantum-gas research \cite{chm17,nye19,mbs13,ghg10,csw18}, the PA rate constants are relatively small at convenient detunings for experiments. A useful figure of merit is the ratio of the number-loss-rate for PA of two atoms in a single lattice site ($\dot{N}_{PA}$) to the off-resonant, single-atom, photon-scattering rate ($R\approx 2\pi s_0\gamma^3_{atomic}/\Delta^2$). The ratio of laser intensity to the atomic transition's saturation intensity ($I_{sat}=40\,$mW/cm$^2$) is indicated by $s_0$.

For two $^{87}$Sr atoms in a single site of an optical lattice, each of mass $m$, $\dot{N}_{PA}\approx 2K_{\nu}\int \,d^3r\,n^2(r)=2K_{\nu}/(2\pi a_{HO}^2)^{3/2}$, where we have assumed the atoms are both in the ground state of a single site in a deep optical lattice in different internal spin states \cite{jzo05}. $a_{HO}=\sqrt{\hbar/m\omega}$ is the harmonic oscillator length for $\omega=\sqrt{4V_0 E_R}/\hbar$ and lattice depth $V_0$.  $E_R=2\pi^2 \hbar^2/(m\lambda_{lat}^2)$ is the recoil energy for lattice laser wavelength $\lambda_{lat}$. For $\lambda_{lat}=1064$\,nm and $V_0=16E_R$, $1/(2\pi a_{HO}^2)^{3/2}=1\times 10^{14}$\,cm$^{-3}$.
For PA at small detuning ($\Delta=-10$\,GHz), $\dot{N}_{PA}\approx 3\times R$. For detuning beyond the wave-function node ($\Delta=-300$\,GHz), $\dot{N}_{PA}\approx 13\times R$, which is more favorable. At $\Delta=-300$\,GHz, a PA laser intensity of 20$I_{sat}$ yields $\dot{N}_{PA}\approx 10/\mathrm{s}$.

In summary, we have measured and characterized  photoassociation resonances up to $\sim 314$ GHz red detuned from the atomic asymptote of the $^1\Sigma^+_u$ $5s^2\,^1$S$_0+5s5p\,^1$P$_1$ molecular potential in fermionic $^{87}$Sr. Similar measurements were made in bosonic $^{86}$Sr and $^{84}$Sr, and a combined fit to the semiclassical LeRoy-Bernstein model allowed determination of the $^1$P$_1$ atomic state lifetime and other spectroscopic parameters. Resonance intensities were compared with predictions from a reflection approximation and the ground-state wavefunction calculated with the best available ground-state potential. We find that, to within experimental uncertainties, the resonance frequencies and intensities are reasonably well-described by theoretical predictions. This work will enable accurate prediction of photoassociative transition frequencies and rates for experiments with the strontium isotopes most commonly used in quantum gas experiments, including experiments with fermionic $^{87}$Sr.





\section*{Acknowledgments}
\label{sec:Acknowledgments}
This work was supported by the National Science Foundation (PHY-1607665), Welch Foundation C-1844 and the AFOSR (FA9550-14-1-0007).

\end{document}